\documentclass[12pt,graphicx,subfigure,axodraw]{article}
\usepackage{amsfonts}
\usepackage{amssymb}
\usepackage[usenames,dvipsnames]{color}
\usepackage{graphicx}

\def\rmuu{\gamma^{\mu}}
\def\rmud{\gamma_{\mu}}
\def\PL{{1-\gamma_5\over 2}}
\def\PR{{1+\gamma_5\over 2}}
\def\sinW2{\sin^2\theta_W}
\def\AEM{\alpha_{EM}}
\def\mul{M_{\tilde{u} L}^2}
\def\mur{M_{\tilde{u} R}^2}
\def\mdl{M_{\tilde{d} L}^2}
\def\mdr{M_{\tilde{d} R}^2}
\def\mz2{M_{z}^2}
\def\c2b{\cos 2\beta}
\def\au{A_u}
\def\ad{A_d}
\def\cob{\cot \beta}
\def\v#1{v_#1}
\def\tb{\tan\beta}
\def\epem{$e^+e^-$}
\def\KK{$K^0$-$\overline{K^0}$}
\def\wi{\omega_i}
\def\xj{\chi_j}
\def\Wmu{W_\mu}
\def\Wnu{W_\nu}
\def\m#1{{\tilde m}_#1}
\def\mH{m_H}
\def\mw#1{{\tilde m}_{\omega #1}}
\def\mx#1{{\tilde m}_{\chi^{0}_#1}}
\def\mc#1{{\tilde m}_{\chi^{+}_#1}}
\def\mwi{{\tilde m}_{\omega i}}
\def\mxi{{\tilde m}_{\chi^{0}_i}}
\def\mci{{\tilde m}_{\chi^{+}_i}}

\def\ch{{\tilde\chi^{+}_1}}
\def\c2{{\tilde\chi^{+}_2}}

\def\tt{{\tilde\theta}}

\def\tp{{\tilde\phi}}

\def\mz{M_z}
\def\sw{\sin\theta_W}
\def\cw{\cos\theta_W}
\def\cb{\cos\beta}
\def\sb{\sin\beta}
\def\rwi{r_{\omega i}}
\def\rxj{r_{\chi j}}
\def\rfp{r_f'}
\def\Kik{K_{ik}}
\def\Fq2{F_{2}(q^2)}
\def\f{\({\cal F}\)}
\def\d1{{\f(\tilde c;\tilde s;\tilde W)+ \f(\tilde c;\tilde \mu;\tilde W)}}
\def\tw{\tan\theta_W}
\def\sec2w{sec^2\theta_W}
\begin{document}
\baselineskip 18pt
\def\today{\ifcase\month\or
 January\or February\or March\or April\or May\or June\or
 July\or August\or September\or October\or November\or December\fi
 \space\number\day, \number\year}
\def\thebibliography#1{\section*{References\markboth
 {References}{References}}\list
 {[\arabic{enumi}]}{\settowidth\labelwidth{[#1]}
 \leftmargin\labelwidth
 \advance\leftmargin\labelsep
 \usecounter{enumi}}
 \def\newblock{\hskip .11em plus .33em minus .07em}
 \sloppy
 \sfcode`\.=1000\relax}
\let\endthebibliography=\endlist
\def\lsim{\ ^<\llap{$_\sim$}\ }
\def\gsim{\ ^>\llap{$_\sim$}\ }
\def\r2{\sqrt 2}
\def\beq{\begin{equation}}
\def\eeq{\end{equation}}
\def\beqn{\begin{eqnarray}}
\def\eeqn{\end{eqnarray}}
\def\rmuu{\gamma^{\mu}}
\def\rmud{\gamma_{\mu}}
\def\PL{{1-\gamma_5\over 2}}
\def\PR{{1+\gamma_5\over 2}}
\def\sinW2{\sin^2\theta_W}
\def\AEM{\alpha_{EM}}
\def\mul{M_{\tilde{u} L}^2}
\def\mur{M_{\tilde{u} R}^2}
\def\mdl{M_{\tilde{d} L}^2}
\def\mdr{M_{\tilde{d} R}^2}
\def\mz2{M_{z}^2}
\def\c2b{\cos 2\beta}
\def\au{A_u}         
\def\ad{A_d}
\def\cob{\cot \beta}
\def\v#1{v_#1}
\def\tb{\tan\beta}
\def\epem{$e^+e^-$}
\def\KK{$K^0$-$\bar{K^0}$}
\def\wi{\omega_i}
\def\xj{\chi_j}
\def\Wmu{W_\mu}
\def\Wnu{W_\nu}
\def\m#1{{\tilde m}_#1}
\def\mH{m_H}
\def\mw#1{{\tilde m}_{\omega #1}}
\def\mx#1{{\tilde m}_{\chi^{0}_#1}}
\def\mc#1{{\tilde m}_{\chi^{+}_#1}}
\def\mwi{{\tilde m}_{\omega i}}
\def\mxi{{\tilde m}_{\chi^{0}_i}}
\def\mci{{\tilde m}_{\chi^{+}_i}}
\def\mz{M_z}
\def\sw{\sin\theta_W}
\def\cw{\cos\theta_W}
\def\cb{\cos\beta}
\def\sb{\sin\beta}
\def\rwi{r_{\omega i}}
\def\rxj{r_{\chi j}}
\def\rfp{r_f'}
\def\Kik{K_{ik}}
\def\Fq2{F_{2}(q^2)}
\def\f{\({\cal F}\)}
\def\d1{{\f(\tilde c;\tilde s;\tilde W)+ \f(\tilde c;\tilde \mu;\tilde W)}}
\def\tw{\tan\theta_W}
\def\sec2w{sec^2\theta_W}
\def\ch{{\tilde\chi^{+}_1}}
\def\c2{{\tilde\chi^{+}_2}}

\def\tt{{\tilde\theta}}

\def\tp{{\tilde\phi}}

\def\mz{M_z}
\def\sw{\sin\theta_W}
\def\cw{\cos\theta_W}
\def\cb{\cos\beta}
\def\sb{\sin\beta}
\def\rwi{r_{\omega i}}
\def\rxj{r_{\chi j}}
\def\rfp{r_f'}
\def\Kik{K_{ik}}
\def\Fq2{F_{2}(q^2)}
\def\f{\({\cal F}\)}
\def\d1{{\f(\tilde c;\tilde s;\tilde W)+ \f(\tilde c;\tilde \mu;\tilde W)}}

\def\b{$\cal{B}(\tau\to\mu \gamma)$~}

\def\tw{\tan\theta_W}
\def\sec2w{sec^2\theta_W}
\newcommand{\pn}[1]{{\color{blue}{#1}}}

\begin{titlepage}

\begin{center}
{\large {\bf 
$\tau\to \mu \gamma$ Decay  in Extensions 
with a Vector Like Generation 
}}\\
\vskip 0.5 true cm
\vspace{2cm}
\renewcommand{\thefootnote}
{\fnsymbol{footnote}}
 Tarek Ibrahim$^{a,b}$\footnote{Email: tarek-ibrahim@alex-sci.edu.eg} 
  and Pran Nath$^{c}$\footnote{Emal: nath@neu.edu}  
\vskip 0.5 true cm
\end{center}

\noindent
{a. Department of  Physics, Faculty of Science,
University of Alexandria,}\\
{ Alexandria, Egypt\footnote{Permanent address.} }\\ 
{b. Department of Physics, Faculty of Science, Beirut Arab University, 
Beirut, Lebanon\footnote{Current address.}} \\
{c. Department of Physics, Northeastern University,
Boston, MA 02115-5000, USA} \\
\vskip 1.0 true cm

\centerline{\bf Abstract}
An analysis is given of the decay $\tau \to \mu+ \gamma$ in MSSM extensions with 
a vector like generation. Here mixing with the  mirrors allows the possibility of this
decay. The analysis is done at one  loop with the exchange of charginos and neutralinos 
and of sleptons and mirror sleptons in the loops. It is shown that a branching ratio 
\b
 in the range $4.4\times 10^{-8} -10^{-9}$ can be gotten which would
be accessible to improved experiment such as at SuperB factories
 for this decay. The effects of CP violation on 
this decay are also analyzed.  \\

\noindent 
Keywords:{~~Lepton flavor change, $\tau \to \mu \gamma$, vector multiplets}\\
PACS numbers:~13.40Em, 12.60.-i, 14.60.Fg

\medskip

\end{titlepage}

\section{Introduction}
  Violation of lepton flavor is an important indicator  of new physics beyond the standard model.
  In the absence of a CKM type matrix in the leptonic sector,  flavor violations can only arise
  due to new physics and thus decays such as $l_i \to l_j \gamma$ ($i\neq j$)  are important probes
  of new physics. We focus here on the decay $\tau\to \mu+\gamma$ on which Babar Collaboration~\cite{Aubert:2009ag}  and Bell Collaboration~\cite{Hayasaka:2007vc}  have put new limits on the branching ratio. Thus 
  The current experimental limit on the branching ratio of this process  from the BaBar
   Collaboration~\cite{Aubert:2009ag}
     based on 
   470fb$^{-1}$ of data and from the Belle Collaboration~\cite{Hayasaka:2007vc}   
   using 535 fb$^{-1}$ of data is 
   
   \beqn
    {\cal B}(\tau \to \mu + \gamma) < 4.4 \times 10^{-8} ~~~~{\rm at ~} 90\% ~{\rm CL} ~~{\rm (BaBar)}\nonumber\\
     {\cal B}(\tau \to \mu + \gamma) < 4.5 \times 10^{-8} ~~~~{\rm at ~} 90\% ~{\rm CL} ~~{\rm (Belle)}   
     \label{1}
   \eeqn
   At the SuperB factories\cite{O'Leary:2010af,Aushev:2010bq,Biagini:2010cc} (for a review see~\cite{Hewett:2012ns})   
       the limit is expected to reach  $ {\cal B}(\tau \to \mu + \gamma)  \sim 1\times 10^{-9}$ as shown in Fig.(1).
      Thus it is of interest to see if theoretical estimates for this branching ratio lie close to the current 
  experimental limits to be detectable in improved experiment.  

   Here we explore this process in the presence of a new vector like generation in an extension of MSSM. 
   Vector like multiplets arise quite naturally in a variety of grand unified models\cite{vectorlike}
   and some of them can escape supermassive mass growth and  can remain light  down to the 
   electroweak scale.  
  Recently an analysis was given of the EDM of the tau in the framework of
  an extension of the minimal supersymmetric standard model with a vector like 
  multiplets~\cite{Ibrahim:2010va}. 
  Specifically mixing of the standard model  leptons  with the mirror leptons, and mixing of  the sleptons
  with mirror sleptons,  were considered and it was  found that such contributions could put the 
  tau  EDM in the detectable range. 
    Here we extend this analysis to investigate the 
  contributions from a vector like lepton multiplet   
   to the flavor changing process $\tau \to \mu + \gamma$.
   This decay is forbidden at the tree level due to  vector current conservation and can only arise
   at the loop level.
  The current work is a logical extension of the previous works where mixings with a vector like
  multiplet and  with mirrors were 
  considered~\cite{Ibrahim:2011im,Ibrahim:2010hv,Ibrahim:2010va,Ibrahim:2008gg,Ibrahim:2009uv}.  
   Implications of additional vector multiplets in other contexts have been 
  explored by many previous authors (see, e.g.,\cite{Babu:2008ge,Liu:2009cc,Martin:2009bg,Graham:2009gy}).
 Several studies already exist on the analysis of $\tau\to \mu\gamma$
  decay~\cite{previous,bhs,Iltan:2001rp,Lavignac:2001vp,Cheung:2001sb,Abada:2008ea,Davidson:2010xv,Moyotl:2012zz}.   However, none of them explore the class of models discussed here. 

  \begin{figure}
\begin{center}
\includegraphics[scale=.4]{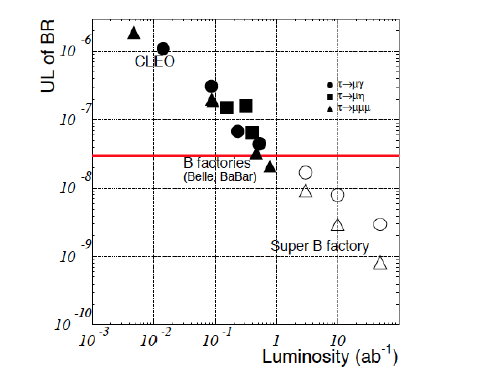}
\caption{A display of the upper limits on the branching ratio  ${\cal B}(\tau^-\to \mu^- \gamma)$ 
(and for $\tau^-\to \mu^-\gamma, \mu^-\mu^+\mu^-$) from the previous experiments and for the
anticipated experiments as a function of the integrated luminosity. Figure is taken from Ref.
\cite{Aushev:2010bq}.}
\end{center}
\label{upper}
\end{figure}
\section{Extension of MSSM with a Vector Multiplet}
We begin with a brief discussion 
on extension of MSSM where we include vector like lepton multiplets since such a  combination 
 is anomaly free. First under $SU(3)_C\times SU(2)_L\times U(1)_Y$ the leptons of the 
 three generations transform as follows
\beqn
\psi_{iL}\equiv 
 \left(\matrix{ \nu_{i L}\cr 
 ~{l}_{iL}}\right)
\sim(1,2,- \frac{1}{2}), l^c_{iL}\sim (1,1,1), ~\nu^c_{i L}\sim (1,1,0), ~i=1,2,3
\label{2}
\eeqn
where the last entry on the right hand side of each $\sim$ is the value of the hypercharge
 $Y$ defined so that $Q=T_3+ Y$.  These leptons have $V-A$ interactions.
We can now add a vector like multiplet where we have a fourth family of leptons with $V-A$ interactions
whose transformations can be gotten from Eq.(\ref{2}) by letting i run from 1-4. 
A vector like lepton multiplet also has  mirrors and so we consider these mirror
leptons which have $V+A$ interactions. Their quantum numbers are
as follows
\beqn
\chi^c\equiv 
 \left(\matrix{ E_{ L}^c\cr 
 N_L^c}\right)
\sim(1,2,\frac{1}{2}), E_{ L}\sim (1,1,-1), N_L\sim (1,1,0).
\label{3}
\eeqn
 The MSSM Higgs  doublets as usual have the quantum numbers  
\beqn
H_1\equiv 
 \left(\matrix{ H_1^1\cr 
 H_1^2}\right)
\sim(1,2,-\frac{1}{2}), ~H_2\equiv 
 \left(\matrix{ H_2^1\cr 
 H_2^2}\right)
\sim(1,2,\frac{1}{2}).
\label{4}
\eeqn

As mentioned already we assume that the vector  multiplet  escapes acquiring mass at the GUT scale and 
remains light down to the electroweak scale. 
As in the analysis  of Ref.\cite{Ibrahim:2010va} 
interesting new physics arises when we consider the 
mixing of the first three generations of leptons with the mirrors of the vector like multiplet.
Actually we will limit ourselves to the second and third generations since only these are relevant for 
the computation of the decay $\tau \to \mu \gamma$. Thus the  superpotential of the model may be written 
in the form
\beqn
W= -\mu \epsilon_{ij} \hat H_1^i \hat H_2^j+\epsilon_{ij}  [f_{1}  \hat H_1^{i} \hat \psi_L ^{j}\hat \tau^c_L
 +f_{1}'  \hat H_2^{j} \hat \psi_L ^{i} \hat \nu^c_{\tau L}
+f_{2}  \hat H_1^{i} \hat \chi^c{^{j}}\hat N_{L}
 +f_{2}'  H_2^{j} \hat \chi^c{^{i}} \hat E_{ L}\nonumber\\
+ h_{1}  H_1^{i} \hat\psi_{\mu L} ^{j}\hat\mu^c_L
 +h_{1}'  H_2^{j} \hat\psi_{\mu L} ^{i} \hat\nu^c_{\mu L}]
+ f_{3} \epsilon_{ij}  \hat\chi^c{^{i}}\hat\psi_L^{j}
 + f_{3}' \epsilon_{ij}  \hat\chi^c{^{i}}\hat\psi_{\mu L}^{j}
 + f_{4}' \hat\mu^c_L \hat E_{ L}  +  f_{5}' \hat\nu^c_{\mu L} \hat N_{L}
 \label{5}
\eeqn
where $\hat\psi_L$ stands for $\hat\psi_{3L}$ and $\hat\psi_{\mu L}$ stands for $\hat\psi_{2L}$.
Here we assume a mixing between the mirror generation and the third lepton generation through
the couplings $f_3$, $f_4$ and $f_5$. We also assume mixing between the mirror generation and the 
second lepton generation through the couplings $f_3'$, $f_4'$ and $f_5'$. 
{\it The above six mass terms are responsible for generating lepton flavor changing process.}
We will focus here on the supersymmetric  sector. Then through the terms $f_3, f_4, f_5, f_3', f_4', f_5'$
 one can have a mixing between
the third generation  and the second generation leptons
which allows the decay of $\tau\to \mu\gamma$  through loop
corrections that include charginos, neutralinos and scalar lepton exchanges
with the photon being emitted by the chargino (see the left diagram of Fig.(2))
or by a charged slepton (see the right diagram of Fig.(2)).
The mass terms for the leptons and  mirrors arise from the term
\beq
{\cal{L}}=-\frac{1}{2}\frac{\partial ^2 W}{\partial{A_i}\partial{A_j}}\psi_ i \psi_ j+H.c.
\label{6}
\eeq
where $\psi$ and $A$ stand for generic two-component fermion and scalar fields.
After spontaneous breaking of the electroweak symmetry, ($<H_1^1>=v_1/\sqrt{2} $ and $<H_2^2>=v_2/\sqrt{2}$),
we have the following set of mass terms written in the 4-component spinor notation
\beqn
-{\cal L}_m =  \left(\matrix{\bar \tau_R & \bar E_{ R} &\bar \mu_R }\right)
 \left(\matrix{f_1 v_1/\sqrt{2} & f_4 & 0\cr
 f_3 & f_2' v_2/\sqrt{2} & f_3'\cr
0 &f_4'& h_1 v_1/\sqrt{2}}\right)\left(\matrix{ \tau_L\cr 
 E_{ L} \cr
\mu_L}\right)+\nonumber\\ 
  + \left(\matrix{\bar \nu_{\tau R} & \bar N_R & \bar \nu_{\mu R} }\right)
 \left(\matrix{f'_1 v_2/\sqrt{2} & f_5 & 0\cr
 -f_3 & f_2 v_1/\sqrt{2} & -f_3'\cr
0&f_5'&h_1' v_2/\sqrt{2}}\right)\left(\matrix{ \nu_{\tau L}\cr 
 N_L \cr
\nu_{\mu L}}\right)  + H.c.
\label{7}
\eeqn 
Here 
the mass matrices are not  Hermitian and one needs
to use bi-unitrary transformations to diagonalize them. Thus we write the linear transformations
\beqn
 \left(\matrix{ \tau_R\cr 
 E_{ R} \cr
\mu_R}\right)=D^{\tau}_R \left(\matrix{ \tau_{1_R}\cr 
 \tau_{2_R}  \cr
\tau_{3_R}}\right),\nonumber\\
\left(\matrix{ \tau_L\cr 
 E_{ L} \cr
\mu_L }\right)=D^{\tau}_L \left(\matrix{ \tau_{1_L}\cr 
 \tau_{2_L} \cr
\tau_{3_L}}\right),
\label{8}
\eeqn
such that
\beq
D^{\tau \dagger}_R 
\left(\matrix{f_1 v_1/\sqrt{2} & f_4 & 0\cr
 f_3 & f_2' v_2/\sqrt{2} & f_3'\cr
0 &f_4'& h_1 v_1/\sqrt{2}}\right)
 D^{\tau}_L=diag(m_{\tau_1},m_{\tau_2},m_{\tau_3}).
 \label{9}
\label{put1}
\eeq
The same holds for the neutrino mass matrix 
\beq
D^{\nu \dagger}_R
 \left(\matrix{f'_1 v_2/\sqrt{2} & f_5 & 0\cr
 -f_3 & f_2 v_1/\sqrt{2} & -f_3'\cr
0&f_5'&h_1' v_2/\sqrt{2}}\right)
 D^{\nu}_L=diag(m_{\nu_1},m_{\nu_2},m_{\nu_2}).
\label{put2}
\label{10}
\eeq
In Eq.(\ref{9})
$\tau_1, \tau_2,\tau_3$  are the mass eigenstates and we identify the tau lepton 
with the eigenstate 1, i.e.,  $\tau=\tau_1$, and we identify $\tau_2$ with a heavy 
mirror eigenstate  with a mass in the hundreds  of GeV and $\tau_3$  is identified as
the muon. Similarly 
$\nu_1, \nu_2, \nu_3$ are the mass eigenstates for the neutrinos, 
where we identify $\nu_1$ as the light tau neutrino, $\nu_2$ as the 
heavier mass eigen state and $\nu_3$ as the muon neutrino.

Next we  consider  the mixings of the charged sleptons and the charged mirror sleptons. 
The mass$^2$ matrix of the slepton - mirror slepton comes from three sources, the F term, the
D term of the potential and soft susy breaking terms.
Using the  superpotential of Eq.(\ref{5}) the mass terms arising from it 
after the breaking of  the electroweak symmetry are given by $ {\cal L}_F$ and $ {\cal L}_D$ 
\beqn
-{\cal L}_F=(m^2_E +|f_3|^2+|f_3'|^2)\tilde E_R \tilde E^*_R +(m^2_N +|f_3|^2+|f_3'|^2)\tilde N_R \tilde N^*_R\nonumber\\
+(m^2_E +|f_4|^2+|f_4'|^2)\tilde E_L \tilde E^*_L
 +(m^2_N +|f_5|^2+|f_5'|^2)\tilde N_L \tilde N^*_L\nonumber\\
+(m^2_{\tau} +|f_4|^2)\tilde \tau_R \tilde \tau^*_R
+(m^2_{\nu_\tau} +|f_5|^2)\tilde \nu_{\tau R} \tilde \nu^*_{\tau R}
+(m^2_{\tau} +|f_3|^2)\tilde \tau_L \tilde \tau^*_L\nonumber\\
+(m^2_{\mu} +|f_4'|^2)\tilde \mu_R \tilde \mu^*_R
+(m^2_{\mu} +|f_3'|^2)\tilde \mu_L \tilde \mu^*_L
+(m^2_{\nu_\tau} 
+|f_3|^2)\tilde \nu_{\tau L} \tilde \nu^*_{\tau L}\nonumber\\
+(m^2_{\nu_\mu} 
+|f_3'|^2)\tilde \nu_{\mu L} \tilde \nu^*_{\mu L}
+(m^2_{\nu_\mu} 
+|f_5'|^2)\tilde \nu_{\mu R} \tilde \nu^*_{\mu R}\nonumber\\
+\{-m_{\tau} \mu^* \tan\beta \tilde \tau_L \tilde \tau^*_R -m_{N} \mu^* \tan\beta \tilde N_L \tilde N^*_R 
-m_{\nu_\tau} \mu^* \cot\beta \tilde \nu_{\tau L} \tilde \nu^*_{\tau R}\nonumber\\
-m_{\mu} \mu^* \tan\beta \tilde \mu_L \tilde \mu^*_R 
-m_{\nu_\mu} \mu^* \cot\beta \tilde \nu_{\mu L} \tilde \nu^*_{\mu R}
\nonumber\\
 -m_{E} \mu^* \cot\beta \tilde E_L \tilde E^*_R 
+(m_E f^*_3 +m_{\tau} f_4) \tilde E_L \tilde \tau^*_L \nonumber\\
+(m_E f_4 +m_{\tau} f^*_3) \tilde E_R \tilde \tau^*_R
+(m_E f'^*_3 +m_{\mu} f_4') \tilde E_L \tilde \mu^*_L\nonumber\\ 
+(m_E f_4' +m_{\mu} f'^*_3) \tilde E_R \tilde \mu^*_R
+(m_{\nu_\tau} f_5 -m_{N} f^*_3) \tilde N_L \tilde \nu^*_{\tau L}\nonumber\\
+(m_{N} f_5 -m_{\nu_\tau} f^*_3) \tilde N_R \tilde \nu^*_{\tau R}
+(m_{\nu_\mu} f'_5 -m_{N} f^*_3) \tilde N_L \tilde \nu^*_{\mu L}\nonumber\\
+(m_{N} f'_5 -m_{\nu_\mu} f'^*_3) \tilde N_R \tilde \nu^*_{\mu R}
+f'_3 f^*_3 \tilde \mu_L \tilde \tau^*_L +f_4 f'^*_4 \tilde \mu_R \tilde \tau^*_R\nonumber\\
+f'_3 f^*_3 \tilde \nu_{\mu L} \tilde \nu_{\tau^*_L} +f_5 f'^*_5 \tilde \nu_{\mu R} \tilde \nu^*_{\tau R}
+H.c. \}.
\label{11}
\eeqn
Similarly the mass terms arising from the D term are given by 
\beqn
-{\cal L}_D=\frac{1}{2} m^2_Z \cos^2\theta_W \cos 2\beta \{\tilde \nu_{\tau L} \tilde \nu^*_{\tau L} -\tilde \tau_L \tilde \tau^*_L 
+\tilde \nu_{\mu L} \tilde \nu^*_{\mu L} -\tilde \mu_L \tilde \mu^*_L
+\tilde E_R \tilde E^*_R -\tilde N_R \tilde N^*_R\}\nonumber\\
+\frac{1}{2} m^2_Z \sin^2\theta_W \cos 2\beta \{\tilde \nu_{\tau L} \tilde \nu^*_{\tau L} +\tilde \tau_L \tilde \tau^*_L 
+\tilde \nu_{\mu L} \tilde \nu^*_{\mu L} +\tilde \mu_L \tilde \mu^*_L \nonumber\\
-\tilde E_R \tilde E^*_R -\tilde N_R \tilde N^*_R +2 \tilde E_L \tilde E^*_L -2 \tilde \tau_R \tilde \tau^*_R
-2 \tilde \mu_R \tilde \mu^*_R
\}.
\label{12}
\eeqn
In addition we have the following set of soft breaking terms  
\beqn
V_{soft}=\tilde M^2_{\tau L} \tilde \psi^{i*}_{\tau L} \tilde \psi^i_{\tau L} 
+\tilde M^2_{\chi} \tilde \chi^{ci*} \tilde \chi^{ci}
+\tilde M^2_{\mu L} \tilde \psi^{i*}_{\mu L} \tilde \psi^i_{\mu L} 
+\tilde M^2_{\nu_\tau} \tilde \nu^{c*}_{\tau L} \tilde \nu^c_{\tau L}
\nonumber\\
 +\tilde M^2_{\nu_\mu} \tilde \nu^{c*}_{\mu L} \tilde \nu^c_{\mu L}
+\tilde M^2_{\tau} \tilde \tau^{c*}_L \tilde \tau^c_L 
+\tilde M^2_{\mu} \tilde \mu^{c*}_L \tilde \mu^c_L 
+\tilde M^2_E \tilde E^*_L \tilde E_L
 + \tilde M^2_N \tilde N^*_L \tilde N_L \nonumber\\
+\epsilon_{ij} \{f_1 A_{\tau} H^i_1 \tilde \psi^j_{\tau L} \tilde \tau^c_L 
-f'_1 A_{\nu_\tau} H^i_2 \tilde \psi ^j_{\tau L} \tilde \nu^c_{\tau L}
+h_1 A_{\mu} H^i_1 \tilde \psi^j_{\mu L} \tilde \mu^c_L 
-h'_1 A_{\nu_\mu} H^i_2 \tilde \psi ^j_{\mu L} \tilde \nu^c_{\mu L}\nonumber\\
+f_2 A_N H^i_1 \tilde \chi^{cj} \tilde N_L
-f'_2 A_E H^i_2 \tilde \chi^{cj} \tilde E_L +H.c.\}
\label{13}
\eeqn
From ${\cal L}_{F,D}$ and
by giving the neutral Higgs their vacuum expectation values in  $V_{soft}$ we can produce the
the mass$^2$  matrix $M^2_{\tilde \tau}$  in the basis $(\tilde  \tau_L, \tilde E_L, \tilde \tau_R, 
\tilde E_R, \tilde \mu_L, \tilde \mu_R)$. We  label the matrix  elements of these as $(M^2_{\tilde \tau})_{ij}= M^2_{ij}$ where
\beqn
M^2_{11}=\tilde M^2_{\tau L} +m^2_{\tau} +|f_3|^2 -m^2_Z cos 2 \beta (\frac{1}{2}-\sin^2\theta_W), \nonumber\\
M^2_{22}=\tilde M^2_E +m^2_{E} +|f_4|^2 +|f'_4|^2 +m^2_Z cos 2 \beta \sin^2\theta_W, \nonumber\\
M^2_{33}=\tilde M^2_{\tau} +m^2_{\tau} +|f_4|^2 -m^2_Z cos 2 \beta \sin^2\theta_W, \nonumber\\
M^2_{44}=\tilde M^2_{\chi} +m^2_{E} +|f_3|^2 +|f'_3|^2 +m^2_Z cos 2 \beta (\frac{1}{2}-\sin^2\theta_W), \nonumber\\
M^2_{55}=\tilde M^2_{\mu L} +m^2_{\mu} +|f'_3|^2 -m^2_Z cos 2 \beta (\frac{1}{2}-\sin^2\theta_W), \nonumber\\
M^2_{66}=\tilde M^2_{\mu} +m^2_{\mu} +|f'_4|^2 -m^2_Z cos 2 \beta \sin^2\theta_W, \nonumber\\
M^2_{12}=M^{2*}_{21}=m_E f^*_3 +m_{\tau} f_4,\nonumber\\
M^2_{13}=M^{2*}_{31}=m_{\tau}(A^*_{\tau} -\mu \tan\beta),\nonumber\\
M^2_{14}=M^{2*}_{41}=0, M^2_{15} =M^{2*}_{51}=f'_3 f^*_3,\nonumber\\
 M^{2*}_{16}= M^{2*}_{61}=0,
M^2_{23}=M^{2*}_{32}=0,\nonumber\\
M^2_{24}=M^{2*}_{42}=m_E(A^*_E -\mu \cot \beta), M^2_{25} = M^{2*}_{52}= m_E f'_3 +m_{\mu} f'^*_4,\nonumber\\
 M^2_{26} =M^{2*}_{62}=0,
M^2_{34}=M^{2*}_{43}=m_E f_4 +m_{\tau} f^*_3, M^2_{35} =M^{2*}_{53} =0, M^2_{36} =M^{2*}_{63}=f_4 f'^*_4\nonumber\\
M^2_{45}=M^{2*}_{54}=0, M^2_{46}=M^{2*}_{64}=m_E f'^*_4 +m_{\mu} f'_3, \nonumber\\
M^2_{56}=M^{2*}_{65}=m_{\mu}(A^*_{\mu} -\mu \tan\beta) 
\label{14}
\eeqn

Here the terms $M^2_{11}, M^2_{13}, M^2_{31}, M^2_{33}$ arise from soft  
breaking in the  sector $\tilde \tau_L, \tilde \tau_R$, 
the terms $M^2_{55}, M^2_{56}, M^2_{65}, M^2_{66}$ arise from soft  
breaking in the  sector $\tilde \mu_L, \tilde \mu_R$,  and 
the terms 
$M^2_{22}, M^2_{24},$  $M^2_{42}, M^2_{44}$ arise from soft  
breaking in the  sector $\tilde E_L, \tilde E_R$. The other terms arise  from mixing between the staus, smuons and 
the mirrors.  We assume that all the masses are of the electroweak size 
so all the terms enter in the mass$^2$ matrix.  We diagonalize this hermitian mass$^2$ matrix  by the
 unitary transformation 
$
 \tilde D^{\tau \dagger} M^2_{\tilde \tau} \tilde D^{\tau} = diag (M^2_{\tilde \tau_1},  
M^2_{\tilde \tau_2}, M^2_{\tilde \tau_3},  M^2_{\tilde \tau_4},  M^2_{\tilde \tau_5},  M^2_{\tilde \tau_6} )$.
There is a  similar mass$^2$  matrix in the sneutrino sector.
In the basis $(\tilde  \nu_{\tau L}, \tilde N_L, \tilde \nu_{\tau R}, \tilde N_R, \tilde  \nu_{\mu L},\tilde \nu_{\mu R} )$ 
we can write the sneutrino mass$^2$ matrix in the form 
$(M^2_{\tilde\nu})_{ij}=m^2_{ij}$ where
\beqn
m^2_{11}=\tilde M^2_{\tau L} +m^2_{\nu_\tau} +|f_3|^2 +\frac{1}{2}m^2_Z cos 2 \beta,  \nonumber\\
m^2_{22}=\tilde M^2_N +m^2_{N} +|f_5|^2 +|f'_5|^2,
~m^2_{33}=\tilde M^2_{\nu_\tau} +m^2_{\nu_\tau} +|f_5|^2,  \nonumber\\
m^2_{44}=\tilde M^2_{\chi} +m^2_{N} +|f_3|^2 +|f'_3|^2 -\frac{1}{2}m^2_Z cos 2 \beta, \nonumber\\
m^2_{55}=\tilde M^2_{\mu L} +m^2_{\nu_\mu} +|f'_3|^2 +\frac{1}{2}m^2_Z cos 2 \beta,  \nonumber\\
m^2_{66}=\tilde M^2_{\nu_\mu} +m^2_{\nu_\mu} +|f'_5|^2,  \nonumber\\
~m^2_{12}=m^{2*}_{21}=m_{\nu_\tau} f_5 -m_{N} f^*_3,\nonumber\\
m^2_{13}=m^{2*}_{31}=m_{\nu_\tau}(A^*_{\nu_\tau} -\mu \cot\beta),
~m^2_{14}=m^{2*}_{41}=0,\nonumber\\
m^2_{14}=m^{2*}_{41}=0, m^2_{15}=m^{2*}_{51}= f'_3 f^*_3, m^2_{16}=m^{2*}_{61}=0,\nonumber\\
~m^2_{23}=m^{2*}_{32}=0,
m^2_{24}=m^{2*}_{42}=m_N(A^*_N -\mu \tan \beta), m^2_{25}=m^{2*}_{52}=-m_N f'_3 +m_{\nu_\mu} f'^*_5,\nonumber\\
m^2_{26}=m^{2*}_{62}=0,
~m^2_{34}=m^{2*}_{43}=m_N f_5 -m_{\nu_\tau} f^*_3,\nonumber\\
m^2_{35}=m^{2*}_{53}=0, m^2_{36}=m^{2*}_{63}=f_5 f'^*_5, m^2_{45}=m^{2*}_{54}=0\nonumber\\
m^2_{46}=m^{2*}_{64}=-m_{\nu_\mu} f'_3 +m_{N} f'^*_5, 
m^2_{56}=m^{2*}_{65}=m_{\nu_\mu}(A^*_{\nu_\mu} -\mu \cot\beta).
\label{15}
\eeqn
As in the charged  slepton sector 
here also the terms $m^2_{11}, m^2_{13}, m^2_{31}, m^2_{33}$ arise from soft  
breaking in the  sector $\tilde \nu_{\tau L}, \tilde \nu_{\tau R}$, 
the terms $m^2_{55}, m^2_{56}, m^2_{65}, m^2_{66}$ arise from soft  
breaking in the  sector $\tilde \nu_{\mu L}, \tilde \nu_{\mu_R}$, and 
  the terms 
$m^2_{22}, m^2_{24},$  $m^2_{42}, m^2_{44}$ arise from soft  
breaking in the  sector $\tilde N_L, \tilde N_R$. The other terms arise  
from mixing between the physical sector and 
the mirror sector.  Again as in the charged lepton sector 
we assume that all the masses are of the electroweak size
so all the terms enter in the mass$^2$ matrix.  This mass$^2$  matrix can be diagonalized  by the
 unitary transformation 
$
 \tilde D^{\nu\dagger} M^2_{\tilde \nu} \tilde D^{\nu} = diag (M^2_{\tilde \nu_1},  
M^2_{\tilde \nu_2}, M^2_{\tilde \nu_3},  M^2_{\tilde \nu_4},M^2_{\tilde \nu_5},  M^2_{\tilde \nu_6} )
$.
The physical tau and neutrino states are $\tau\equiv \tau_1, \nu_{\tau}\equiv \nu_1$,
and the states $\tau_2, \nu_2$ are heavy states with mostly mirror particle content. 
Similarly $\mu\equiv \tau_3, \nu_{\mu}\equiv \nu_3$.
The states $\tilde \tau_i, \tilde \nu_i; ~i=1-6$ are the slepton and sneutrino 
mass eigenstates.
 
\section{Interactions of Charginos and Neutralinos} 

The chargino exchange contribution to the decay of  the tau  into a muon and a photon arises through 
the left loop diagram of Fig.(2).   The relevant part of Lagrangian that 
 generates this contribution is given by 
\beqn
-{\cal{L}}_{\tau-\tilde{\nu}-\chi^+}=
\sum_{\alpha =1}^3\sum_{i=1}^2\sum_{j=1}^6  
\bar{\tau}_{\alpha}[C^L_{\alpha ij} P_L+ 
 C^R_{\alpha ij} P_R] 
\tilde{\chi^c}_i \tilde{\nu}_j +H.c.
\label{16}
\eeqn
where
\beqn
C^L_{\alpha ij}=g[-\kappa_{\tau} U^*_{i2} D^{\tau *}_{R _{1\alpha}}\tilde{D}^{\nu}_{1j}
-\kappa_{\mu} U^*_{i2} D^{\tau *}_{R _{3\alpha}}\tilde{D}^{\nu}_{5j}\nonumber\\
+ U^*_{i1} D^{\tau *}_{R _{2\alpha}}\tilde{D}^{\nu}_{4j}
-\kappa_{N} U^*_{i2} D^{\tau *}_{R _{2\alpha}}\tilde{D}^{\nu}_{2j}],
\nonumber\\
C^R_{\alpha ij}=g[-\kappa_{\nu_{\tau}} V_{i2} D^{\tau *}_{L _{1\alpha}}\tilde{D}^{\nu}_{3j}
-\kappa_{\nu_{\mu}} V_{i2} D^{\tau *}_{L _{3\alpha}}\tilde{D}^{\nu}_{6j}\nonumber\\
+ V_{i1} D^{\tau *}_{L _{1\alpha}}\tilde{D}^{\nu}_{1j}
+ V_{i1} D^{\tau *}_{L _{3\alpha}}\tilde{D}^{\nu}_{5j}
-\kappa_{E} V_{i2} D^{\tau *}_{L _{2\alpha}}\tilde{D}^{\nu}_{4j}],
\label{17}
\eeqn
where $\tilde{D}^{\nu}$ is the diagonalizing matrix of the scalar $6\times 6$ mass$^2$ matrix 
for the scalar neutrino as defined 
above.
  $\kappa_N, \kappa_{\tau}$  etc that enter in the equation above  are defined by 
 \beqn
(\kappa_N, \kappa_{\tau}, \kappa_{\mu})
=\frac{(m_N, m_{\tau}, m_{\mu})}{\sqrt{2} M_W \cos\beta},~
(\kappa_{E}, \kappa_{\nu})    =\frac{(m_{E}, m_{\nu})}{\sqrt{2} M_W \sin\beta}.
\label{18}
\eeqn
In Eq.(\ref{17}) 
 $U$ and $V$ are the matrices  that  diagonalize the chargino mass matrix $M_C$ 
  so that 
\beq
U^* M_C V^{-1}= diag (m_{\tilde{\chi_1}}^+,m_{\tilde{\chi_2}}^+).
\label{19}
\eeq
\begin{figure}
\begin{center}
\includegraphics[scale=.6]{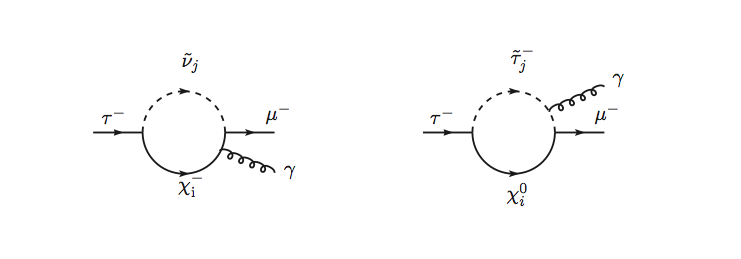}
\caption{\label{taumugamma} 
The diagrams that allow decay of the $\tau$ into $\mu+\gamma$ via
supersymmetric loops involving the chargino and the sneutrino (left) and the neutralino and the stau
(right) with emission of the photon from the charged particle inside the loop.}
\end{center}
\label{fig1}
\end{figure}

The neutralino exchange contribution to the tau decay arises through 
the right loop diagram  of Fig. (2).  
The relevant part of 
Lagrangian that generates this contribution is given by 
\beqn
-{\cal{L}}_{\tau-\tilde{\tau}-\chi^0}=
\sum_{\alpha=1}^3\sum_{i=1}^4\sum_{j=1}^6  
\bar{\tau}_{\alpha}[C'^L_{\alpha ij} P_L+ 
C'^R_{\alpha ij} P_R] 
\tilde{\chi^0}_i \tilde{\tau}_j +H.c.,
\label{20}
\eeqn
where  as stated earlier $\tau=\tau_1$ and $\mu=\tau_3$. In Eq.(\ref{20}) $C_L'$ and $C_R'$ are  defined
by

\beqn
C'^L_{\alpha ij}=\sqrt{2}[\alpha_{\tau i}D^{\tau *}_{R_{1\alpha}} \tilde{D}^{\tau}_{1j}
-\delta_{E i}D^{\tau *}_{R_{2\alpha}} \tilde{D}^{\tau}_{2j}
-\gamma_{\tau i}D^{\tau *}_{R_{1\alpha}} \tilde{D}^{\tau}_{3j}\nonumber\\
+\beta_{E i}D^{\tau *}_{R_{2\alpha}} \tilde{D}^{\tau}_{4j}
+\alpha_{\mu i}D^{\tau *}_{R_{3\alpha}} \tilde{D}^{\tau}_{5j}
-\gamma_{\mu i}D^{\tau *}_{R_{3\alpha}} \tilde{D}^{\tau}_{6j}],
\nonumber\\
C'^R_{\alpha ij}=\sqrt{2}[\beta_{\tau i}D^{\tau *}_{L_{1\alpha}} \tilde{D}^{\tau}_{1j}
-\gamma_{E i}D^{\tau *}_{L_{2\alpha}} \tilde{D}^{\tau}_{2j}
-\delta_{\tau i}D^{\tau *}_{L_{1\alpha}} \tilde{D}^{\tau}_{3j}\nonumber\\
+\alpha_{E i}D^{\tau *}_{L_{2\alpha}} \tilde{D}^{\tau}_{4j}
+\beta_{\mu i}D^{\tau *}_{L_{3\alpha}} \tilde{D}^{\tau}_{5j}
-\delta_{\mu i}D^{\tau *}_{L_{3\alpha}} \tilde{D}^{\tau}_{6j}],
\label{21}
\eeqn
where  $\tilde{D}^{\tau}$ is the diagonalizing matrix of the $6\times 6$ slepton mass$^2$ matrix.
\beqn\label{alphabk}
\alpha_{E_{ j}} =\frac{g m_{E} X^*_{4j}}{2m_W\sin\beta},~~
\beta_{E_{ j}}=eX_{1j}^{'} +\frac{g}{\cos\theta_W} X_{2j}^{'}
(\frac{1}{2}-\sin^2\theta_W),\nonumber\\
\gamma_{E_{ j}}=e X_{1j}^{'*}-\frac{g\sin^2\theta_W}{\cos\theta_W}
X_{2j}^{*'},
~~ \delta_{E_{ j}}=-\frac{g m_{E} X_{4j}}{2m_W \sin\beta}, 
\label{22}
\eeqn
and
\beqn
\alpha_{\tau j} =\frac{g m_{\tau} X_{3j}}{2m_W\cos\beta},~~
\alpha_{\mu j} =\frac{g m_{\mu} X_{3j}}{2m_W\cos\beta},~~
\beta_{\tau j}=\beta_{\mu j}=-eX_{1j}^{'*} +\frac{g}{\cos\theta_W} X_{2j}^{'*}
(-\frac{1}{2}+\sin^2\theta_W),\nonumber\\
\gamma_{\tau j}=\gamma_{\mu j}=-e X_{1j}'+\frac{g\sin^2\theta_W}{\cos\theta_W}
X_{2j}',
~~ \delta_{\tau j}=-\frac{g m_{\tau} X^*_{3j}}{2m_W \cos\beta},
~~ \delta_{\mu j}=-\frac{g m_{\mu} X^*_{3j}}{2m_W \cos\beta},
\label{23}
\eeqn
where
\beqn
X'_{1j}= (X_{1j}\cos\theta_W + X_{2j} \sin\theta_W), 
~X'_{2j}=  (-X_{1j}\sin\theta_W + X_{2j} \cos\theta_W), 
\label{24}
\eeqn
and where the matrix $X$ diagonlizes the neutralino mass matrix so that
\beq
X^T M_{\tilde{\chi}^0} X=diag(m_{{\chi^0}_1}, m_{{\chi^0}_2}, m_{{\chi^0}_3}, m_{{\chi^0}_4}).
\label{25}
\eeq

\section{The analysis of $\tau \rightarrow \mu + \gamma$  Decay Width }

The decay $\tau \rightarrow \mu + \gamma$
is induced by one-loop electric and 
magnetic  transition dipole moments,  which arise
from the diagrams of Fig.(2). In the dipole moment loop, the incoming muon is replaced by a tau lepton.
For an incoming tau of momentum $p$ and a resulting muon of momentum $p'$, we define the amplitude
\beq
<\mu (p') | J_{\alpha} | \tau (p)> = \bar{u}_{\mu} (p') \Gamma_{\alpha} u_{\tau} (p) 
\label{26}
\eeq
where
\beq
\Gamma_{\alpha} (q) =\frac{F^{\tau \mu}_2 (q) i \sigma_{\alpha \beta} q^{\beta}}{m_{\tau} +m_{\mu}}
+\frac{F^{\tau \mu}_3 (q)  \sigma_{\alpha \beta} \gamma_5 q^{\beta}}{m_{\tau} +m_{\mu}}+.....
\label{27}
\eeq
with $q = p' -p$ and where $m_f$ denotes the mass of the fermion $f$.
The branching ratio of $\tau \rightarrow \mu + \gamma$ is given by
\beqn
   {\cal B} (\tau \rightarrow \mu + \gamma) =\frac{24 \pi^2}{5 G^2_F m^2_{\tau} (m_{\tau}+m_{\mu})^2} \{|F^{\tau \mu}_2 (0)|^2
+|F^{\tau \mu}_3 (0)|^2 \}
\label{28}
\eeqn
where the form factors $F^{\tau \mu}_2$ and $F^{\tau \mu}_3$ arise from the chargino and the 
neutralino contributions as follows
\beqn
F^{\tau \mu}_2 (0) = F^{\tau \mu}_{2 \chi^+} + F^{\tau \mu}_{2 \chi^0} \nonumber\\
F^{\tau \mu}_3 (0) = F^{\tau \mu}_{3 \chi^+} + F^{\tau \mu}_{3 \chi^0} 
\label{29}
\eeqn

\noindent
The chargino contribution  $F^{\tau \mu}_{2 \chi^+}$ is given by
\beqn
F^{\tau \mu}_{2 \chi^+}=\sum_{i=1}^2 \sum_{j=1}^6 [ \frac{m_{\tau}(m_{\tau} +m_{\mu})}{64 \pi^2 m^2_{\tilde{\chi_i}^+}}
\{C^L_{3ij} C^{L*}_{1ij} + C^R_{3ij} C^{R*}_{1ij} \} F_1 (\frac{M^2_{\tilde{\nu_j}}}{m^2_{\tilde{\chi_i}^+}})\nonumber\\
+ \frac{(m_{\tau} +m_{\mu})}{64 \pi^2 m_{\tilde{\chi_i}^+}}
\{C^L_{3ij} C^{R*}_{1ij} + C^R_{3ij} C^{L*}_{1ij} \} F_2 (\frac{M^2_{\tilde{\nu_j}}}{m^2_{\tilde{\chi_i}^+}})]
\label{30}
\eeqn
where
\beq
F_1(x)= \frac{1}{3(x-1)^4} \{-2 x^3 -3x^2 +6x -1 +6x^2 ln x \}
\label{31}
\eeq
and 
\beq
F_2(x)= \frac{1}{(x-1)^3} \{3x^2 -4x +1 -2x^2 ln x \}
\label{32}
\eeq

\noindent
The neutralino contribution  $F^{\tau \mu}_{2 \chi^0}$ is given by
\beqn
F^{\tau \mu}_{2 \chi^0}= \sum_{i=1}^4 \sum_{j=1}^6 [\frac{-m_{\tau}(m_{\tau} +m_{\mu})}{192 \pi^2 m^2_{\tilde{\chi_i}^0}}
\{C'^L_{3ij} C'^{L*}_{1ij} + C'^R_{3ij} C'^{R*}_{1ij} \} F_3 (\frac{M^2_{\tilde{\tau_j}}}{m^2_{\tilde{\chi_i}^0}})\nonumber\\
- \frac{(m_{\tau} +m_{\mu})}{64 \pi^2 m_{\tilde{\chi_i}^0}}
\{C'^L_{3ij} C'^{R*}_{1ij} + C'^R_{3ij} C'^{L*}_{1ij} \} F_4 (\frac{M^2_{\tilde{\tau_j}}}{m^2_{\tilde{\chi_i}^0}})]
\label{33}
\eeqn
where
\beq
F_3(x)= \frac{1}{(x-1)^4} \{- x^3 +6x^2 -3x -2 -6x ln x \}
\label{34}
\eeq
and 
\beq
F_4(x)=\frac{1}{(x-1)^3} \{-x^2 +1 +2x ln x \}
\label{35}
\eeq

\noindent
The chargino contribution  $F^{\tau \mu}_{3 \chi^+}$ is given by
\beqn
F^{\tau \mu}_{3 \chi^+}= \sum_{i=1}^2 \sum_{j=1}^6 \frac{(m_{\tau} +m_{\mu})m_{\tilde{\chi_i}^+} }{32 \pi^2 M^2_{\tilde{\nu_j}}}
\{ C^L_{3ij} C^{R*}_{1ij} - C^R_{3ij} C^{L*}_{1ij} \} 
F_5(\frac{m^2_{\tilde{\chi_i}^+}}{M^2_{\tilde{\nu_j}}})
\label{36}
\eeqn
where
\beq
F_5(x)= \frac{1}{2(x-1)^2} \{-x +3 + \frac{2ln x}{1-x} \}
\label{37}
\eeq

\noindent
The neutralino contribution  $F^{\tau \mu}_{3 \chi^0}$ is given by
\beqn
F^{\tau \mu}_{3 \chi^0}= \sum_{i=1}^4 \sum_{j=1}^6 \frac{(m_{\tau} +m_{\mu})m_{\tilde{\chi_i}^0} }{32 \pi^2 M^2_{\tilde{\tau_j}}}
\{ C'^L_{3ij} C'^{R*}_{1ij} - C'^R_{3ij} C'^{L*}_{1ij} \} 
F_6(\frac{m^2_{\tilde{\chi_i}^0}}{M^2_{\tilde{\tau_j}}})
\label{38}
\eeqn
where
\beq
F_6(x)=  \frac{1}{2(x-1)^2} \{x +1 + \frac{2 x ln x}{1-x} \}
\label{39}
\eeq
\begin{figure}
\vspace{-2cm}
    \centering
    \begin{minipage}[b]{.45\linewidth}
        \centering
        \includegraphics[scale=.25]{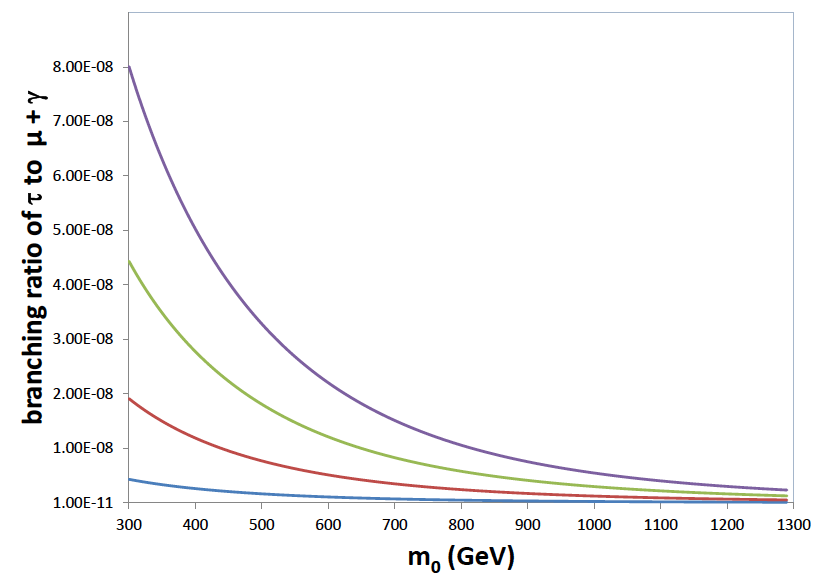}        
    \end{minipage}%
    \hfill%
    \begin{minipage}[b]{.45\linewidth}
    \centering
        \centering
  \includegraphics[scale=.25]{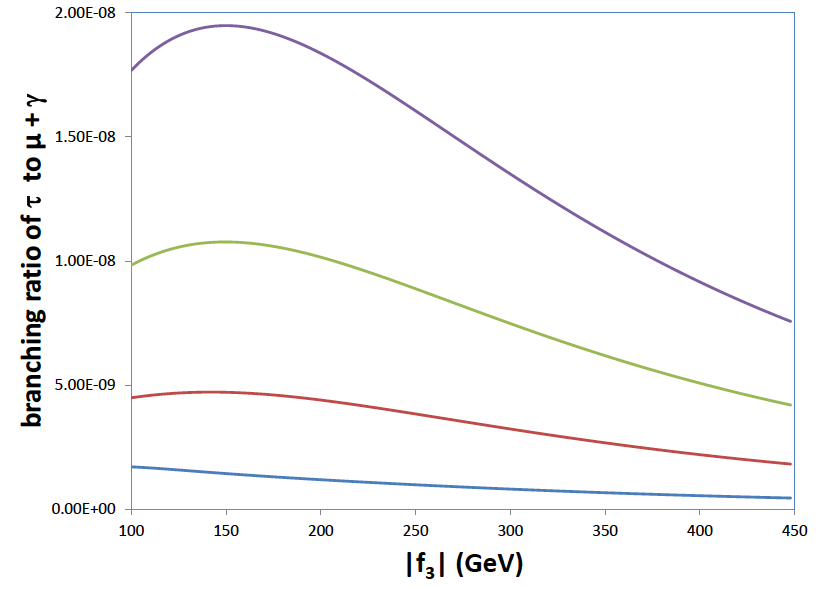}  
    \end{minipage}\\[-7pt]
    \begin{minipage}[t]{.45\linewidth}
        \caption{
        An exhibition of the dependence of 
        \b 
        on $m_0$ when 
$m_N=120$, $m_E=150$,  $|f_3|=|f'_3|=$90, $|f_4|=|f'_4|=$100,
$|f_5|=|f'_5|=$80, $|A_0|=$150, $\tilde{m}_1=50$, $\tilde{m}_2=100$, $\mu=150$, $\chi_3=\chi'_3=$0.6, $\chi_4=\chi'_4=$0.4, $\chi_5=\chi'_5=$0.6, $\alpha_E=$0.5,  $\alpha_N=$0.8,  
 and  $\tan \beta=$5, 10, 15, 20 (in ascending order at $m_0=300$).
 Here and in Figs.(4-7) masses are in GeV and angles are in rad. 
 }        
       \end{minipage}%
    \hfill%
    \begin{minipage}[t]{.45\linewidth}
        \caption{
        An exhibition of the dependence of \b
        on $|f_3|$ when $m_0=$900,
$m_N=150$, $m_E=180$,  $|f'_3|=$100, $|f_4|=|f'_4|=$100,
$|f_5|=|f'_5|=$70, $|A_0|=$100, $\tilde{m}_1=50$, $\tilde{m}_2=100$, $\mu=150$, $\chi_3=\chi'_3=$0.6, $\chi_4=\chi'_4=$0.4, $\chi_5=\chi'_5=$0.6, $\alpha_E=$0.5,  $\alpha_N=$0.8,  
 and  $\tan \beta=$5, 10, 15, 20 (in ascending order at $|f_3|=100$.) }
    \end{minipage}%
\end{figure}

\section{Estimate of size of  \b }

In this section we give a numerical analysis of \b for the model where we include a leptonic vector 
multiplet.  As discussed in the previous sections the flavor changing processes arise from the mixings
between the standard model leptons and the mirrors in the vector multiplet. 
The mixing matrices between leptons and mirrors are diagonalized using bi-unitary 
transformations with
matrices $D^{\tau}_R$ and $D^{\tau}_L$.
The input parameters for this sector of the parameter space are
$m_{\tau},m_E,m_{\mu}$, $f_3$, $f_4$, $f'_3$, $f'_4$ where
 $f_3$, $f_4$, $f'_3$ and $f'_4$ are complex masses with CP violating phases $\chi_3$, $\chi_4$, $\chi'_3$
$\chi'_4$.
 For the slepton mass$^2$ matrices we need the extra input parameters of the susy breaking sector,
$\tilde{M}{\tau_L}, \tilde{M}_E,\tilde{M}_{\tau},\tilde{M}_{\chi},\tilde{M}_{\mu_L},\tilde{M}_{\mu},
A_{\tau}, A_E, A_{\mu},A_{N}, \mu, \tan\beta$.
For the sneutrino mass$^2$ matrices we have more input parameters,
$\tilde{M}_{N}, \tilde{M}_{\nu_{\tau}},\tilde{M}_{\nu_{\mu}},A_{\nu_{\mu}},A_N, A_{\nu_e},$
$m_{N}, f_5, f'_5.$
The chargino and neutralino sectors need the extra two parameters
$\tilde{m}_1, \tilde{m}_2$.
In the analysis we will include phases since dipole moments are sensitive to phases (for a review
see ~\cite{Ibrahim:2007fb}).
Here for simplicity 
we assume that the only parameters that are complex in the above matrix elements are 
$A_E$, $A_N$, $A_{\tau}$, $A_{\mu}$, $A_{\nu}$, $f_5$ and $f'_5$ which have the phases 
$\alpha_E$, $\alpha_N$, $\alpha_{\tau}$, $\alpha_{\mu}$, $\alpha_{\nu}$, $\chi_5$ and $\chi'_5$.
To simplify the analysis we set $\alpha_{\nu}=\alpha_{\mu}=\alpha_{\tau}=0$.
Thus the CP violating phases that would play a role in this analysis are
\beq
\chi_3, \chi_4, \chi_5, \chi'_3, \chi'_4, \chi'_5,  \alpha_E, \alpha_N.
\label{40}
\eeq
With the above  in mind, the electric dipole moments of the electron, the neutron and of the Hg atom vanish and we do not 
need to worry about  them satisfying their upper limit constraints.
To reduce the number of input parameters we assume equality of 
the scalar masses and of the trilinear couplings so that
$\tilde{M}_a =m_0, a={\tau_L}, E, \tau, \chi, \nu,\mu, {\mu_L}, N$ and $|A_i|=|A_0|$, $i=E, N, \tau, \nu, \mu$. 
 \begin{figure}
    \centering
    \begin{minipage}[b]{.45\linewidth}
        \centering
        \includegraphics[scale=.29]{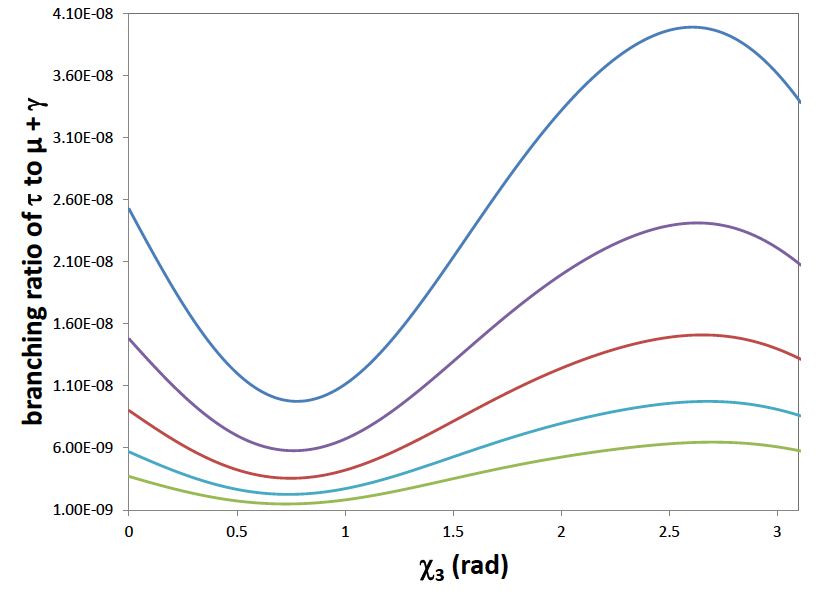}        
    \end{minipage}%
    \hfill%
    \begin{minipage}[b]{.45\linewidth}
    \centering
        \centering
  \includegraphics[scale=.29]{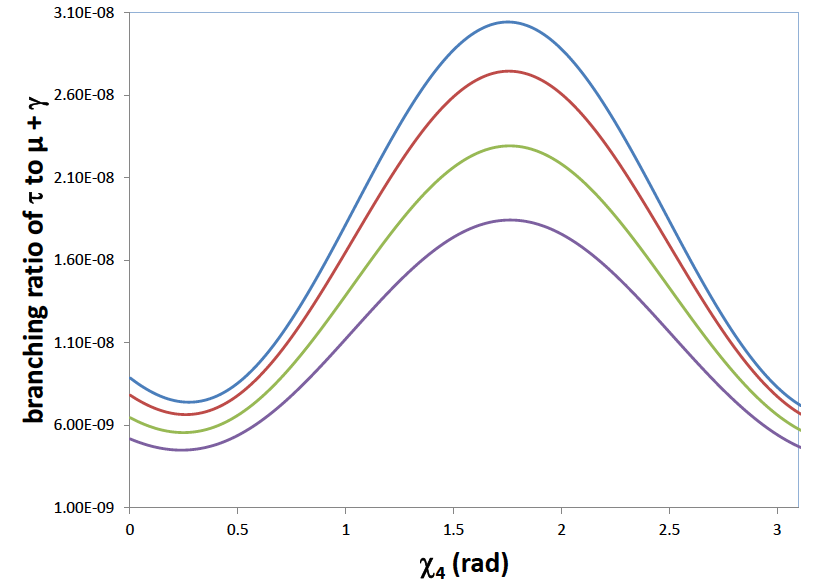}  
    \end{minipage}\\[-7pt]
    \begin{minipage}[t]{.45\linewidth}
        \caption{
       An exhibition of the dependence of \b
       on $\chi_3$ when $\tan\beta=$10,
$m_N=170$, $m_E=200$,  $|f_3|=|f'_3|=$250, $|f_4|=|f'_4|=$400,
$|f_5|=|f'_5|=$90, $|A_0|=$130, $\tilde{m}_1=90$, $\tilde{m}_2=80$, $\mu=120$, $\chi'_3=$0.8, $\chi_4=\chi'_4=$0.9, $\chi_5=\chi'_5=$1.6, $\alpha_E=$1.0,  $\alpha_N=$0.9,  
 and  $m_0=$900, 800, 700, 600, 500  (in ascending order at $\chi_3=0.0$.)}
       \end{minipage}%
    \hfill%
    \begin{minipage}[t]{.45\linewidth}
        \caption{
  An exhibition of the dependence of \b
  on $\chi_4$ when $m_0=$800, $\tan\beta=$15,
$m_N=160$, $m_E=220$,  $|f'_3|=$150, $|f_4|=|f'_4|=$200,
$|f_5|=|f'_5|=$100, $|A_0|=$160, $\tilde{m}_1=100$, $\tilde{m}_2=90$, $\mu=150$, $\chi_3=\chi'_3=$0.6, $\chi'_4=$0.8, $\chi_5=\chi'_5=$1.0, $\alpha_E=$.4,  $\alpha_N=$0.8,  
 and  $|f_3|=$300, 250, 200, 150  (in ascending order at $\chi_4=0.0$.) }
    \end{minipage}%
\end{figure}

Fig.(3) gives an analysis of \b  as a function of  $m_0$ for values of $\tan\beta=5,10,15, 20$ with other inputs as given in the caption of Fig.(3).
The branching ratio depends on the chargino and neutralino exchange contributions to $F_2$ and $F_3$  defined in Eq.(\ref{29})
which depend on $m_0$ through the slepton masses that enter the loops.
Fig.(3) exhibits a sharp dependence on $\tan\beta$ which enters $F_2$ and $F_3$ also via the slepton masses as well as through
the chargino and neutralino mass matrices. Further, the couplings $C^{L,R}$ and $C'^{L,R}$ are also affected by variations in 
$m_0$ and $\tan\beta$. The analysis of Fig.(3) shows that there is a significant part of the parameter space where \b lies in the 
range $O(10^{-8})$ consistent with the  upper limit of Eq.(\ref{1}).
Fig.(4) gives an analysis of \b as a function of  $|f_3|$,  where $f_3$ is an off diagonal term in the mass matrix of
    Eq.(\ref{7}), 
 for $\tan\beta$ values as in Fig.(\ref{3}) and the other inputs are as given in the caption of Fig.(4). As in Fig.(3) one finds a sharp dependence
 on $\tan\beta$.  
 This dependence  of $|f_3|$ arises since  
 it enters in the  matrix  elements diagonalizing matrices  $D^{\tau}_{L,R}$ and this way it affects both 
 chargino and neutralino exchange  contributions. The entire parameter space exhibited in this figure is consistent with the
 upper limits of Eq.(\ref{1}). 

\begin{figure}
\vspace{-2cm}
\begin{center}
\includegraphics[scale=.29]{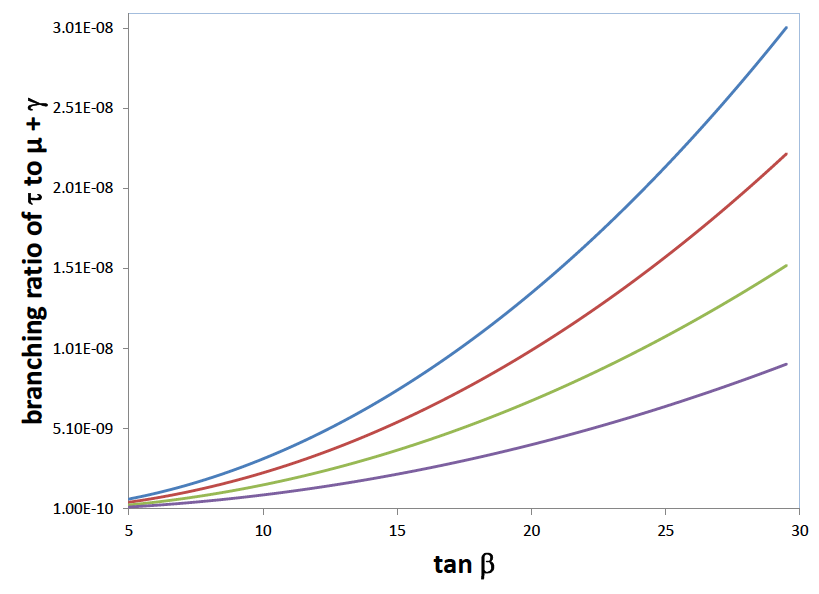}
\caption{
 An exhibition of the dependence of \b
 on $\tan\beta$ when $m_0=$700, 
$m_N=200$, $m_E=300$,  $|f_3|=|f'_3|=$180, $|f_4|=|f'_4|=$100,
$|f_5|=|f'_5|=$150, $|A_0|=$360, $\tilde{m}_1=120$, $\tilde{m}_2=80$, $\mu=140$, $\chi_3=\chi'_3=$0.7, $\chi_4=\chi'_4=$0.9, $\chi_5=\chi'_5=$.6, $\alpha_E=$.9,  $\alpha_N=$0.4,  
 and  $\chi_3=$1.2, 0.8, 0.5, 0.1  (in ascending order at $\tan\beta=30$.)}
\end{center}
\label{fig2}
\end{figure}
 We discuss now the effect of CP phases on ${\cal B}(\tau\to \mu\gamma)$. As mentioned above the phases of Eq.(\ref{40}) 
 have no effect on the EDMs of the electron,  on  the EDM of the neutron
or on EDM of  the Hg atom and these phases only affect phenomena related to the second and the third generation leptons. 
  Fig.(5) gives a display of \b  as a function of $\chi_3$ for values of $m_0=900,800,700,600,500$ GeV 
  (in ascending order)  when
  $\tan\beta =10$ and the other inputs are as shown in the caption of Fig.(5).  Here one finds that \b has a significant  dependence
  on $\chi_3$. Thus, for instance, for the case $m_0=500$ GeV (top curve) one finds that \b can vary in the range
  $(1\times 10^{-8}- 4\times 10^{-8})$ as $\chi_3$ varies in the range $(0,\pi)$. Again \b displayed in this analysis is consistent with  the upper limit of Eq.(\ref{1}) over 
  the entire range of parameters exhibited.  
  
  Another analysis on the dependence of \b on CP phases is exhibited in Fig.(6) where a plot of \b as a function of $\chi_4$ is given 
  for the case when $|f_3|=(300, 250,$ $200,$ $150)$ GeV (in ascending order), $\tan\beta=15$ and other inputs are as given in the caption of Fig.(6).  Again a very
  significant variation in \b is seen as  $\chi_4$ varies in the range $(0,\pi)$. Specifically one finds that for the case $|f_3|=150$, \b  varies
  in the range $(8\times 10^{-9}-3\times 10^{-8})$. Further, over the entire parameter space 
  analysed in Fig.(6) 
  \b is consistent with the upper
  limit of Eq.(\ref{1}). 
  Finally, in Fig.(7) we exhibit the dependence of \b on $\tan\beta$ when $\chi_3= 1.2, 0.8, 0.5, 0.1$ (in ascending order) 
 with other parameters as defined in the caption of Fig.(7).  A  sharp dependence of \b on $\tan\beta$ can
  be seen. Specifically one finds that for the case $\chi_3=0.1$ (the top curve)  \b varies in the range $(1\times 10^{-10}-3\times 10^{-8})$ which is 
  more than an order of magnitude variation as $\tan\beta$ varies in the range (5-30).  
 
 In summary in the analyses presented in Fig.(3-7), one finds that \b can be quite large often lying just below the current experimental limits  which implies that this part of the parameter space will be accessible to future experiments,  specifically SuperB factories  which can probe \b as low as $10^{-9}$.  We note that the  flavor changing
 interactions of Eq.(\ref{5}) also contribute to the muon anomalous magnetic moment $g_{\mu}-2$ which is very
 precisely determined experimentally.   This   can come
 about by the exchange of a tau and a photon in the loop but since each vertex is one loop order, the 
 contribution is three loop order which would be tiny compared to other standard model electroweak
 contributions.

\section{Conclusion}
Lepton flavor changing processes provide an important window to new physics beyond the standard 
model.  In this work we have analyzed the decay $\tau \to \mu + \gamma$ in  extensions of the 
MSSM with  vector like leptonic multiplets which are anomaly free. Specifically we consider mixings between
the standard model generations of leptons with the mirror leptons in the vector multiplet.
{\it It is because of these mixings which are parametrized by $f_3, f_4, f_5$ and $f_3', f_4', f_5'$ as
defined in Eq.(\ref{5}) that lepton flavor violations appear.}
 We focus
on the supersymmetric sector and compute 
 contributions to this process arising 
 from diagrams with  exchange of charginos and sneutrinos  in the loop and with the exchange of neutralinos and staus in the
 loop. These loops  do not preserve lepton flavor. 
A full analytic analysis of these loops was given which constitute the main result of this work.
A numerical analysis was also carried out and it is found that there exists a significant part of the 
parameter space where one can have the branching ratio for this process in the range $4.4\times 10^{-8} - 10^{-9}$,
where $4.4\times 10^{-8}$  at 90\% CL is the upper limit from BaBar (see Eq.(\ref{1}))
and the lower limit is the sensitivity that the SuperB factories will achieve.
Thus it is very likely that improved experiment with a better  sensitivity may be
able to probe this class of models. \\

\noindent
{\em Acknowledgments}:  
This research is  supported in part by  NSF grant PHY-0757959 and PHY-0704067.


\begin{thebibliography}{999}

\bibitem{Aubert:2009ag} 
  B.~Aubert {\it et al.}  [BABAR Collaboration],
  ``Searches for Lepton Flavor Violation in the Decays tau+-$\to$  e+- gamma and tau+- $\to$  mu+- gamma,''
  Phys.\ Rev.\ Lett.\  {\bf 104}, 021802 (2010)
  [arXiv:0908.2381 [hep-ex]].

\bibitem{Hayasaka:2007vc} 
  K.~Hayasaka {\it et al.}  [Belle Collaboration],
  ``New search for tau $\to$ mu gamma and tau $\to$ e gamma decays at Belle,''
  Phys.\ Lett.\ B {\bf 666}, 16 (2008)
  [arXiv:0705.0650 [hep-ex]].

\bibitem{O'Leary:2010af} 
  B.~O'Leary {\it et al.}  [SuperB Collaboration],
  ``SuperB Progress Reports -- Physics,''
  arXiv:1008.1541 [hep-ex].

\bibitem{Aushev:2010bq} 
  T.~Aushev, W.~Bartel, A.~Bondar, J.~Brodzicka, T.~E.~Browder, P.~Chang, Y.~Chao and K.~F.~Chen {\it et al.},
  ``Physics at Super B Factory,''
  arXiv:1002.5012 [hep-ex].

\bibitem{Biagini:2010cc} 
  M.~E.~Biagini {\it et al.}  [SuperB Collaboration],
  ``SuperB Progress Reports: The Collider,''
  arXiv:1009.6178 [physics.acc-ph].

\bibitem{Hewett:2012ns} 
  J.~L.~Hewett, H.~Weerts, R.~Brock, J.~N.~Butler, B.~C.~K.~Casey, J.~Collar, A.~de Govea and R.~Essig {\it et al.},
  ``Fundamental Physics at the Intensity Frontier,''
  arXiv:1205.2671 [hep-ex].


\bibitem{vectorlike}
  H.~Georgi,
  ``Towards A Grand Unified Theory Of Flavor,''
  Nucl.\ Phys.\  B {\bf 156}, 126 (1979);
  F.~Wilczek and A.~Zee,
  ``Families From Spinors,''
  Phys.\ Rev.\  D {\bf 25}, 553 (1982);
J. Maalampi, J.T. Peltoniemi, and M. Roos, PLB 220, 441(1989);
  J.~Maalampi and M.~Roos,
  ``Physics Of Mirror Fermions,''
  Phys.\ Rept.\  {\bf 186}, 53 (1990);
  K.~S.~Babu, I.~Gogoladze, P.~Nath and R.~M.~Syed,
   ``A Unified framework for symmetry breaking in SO(10),''
  Phys.\ Rev.\ D {\bf 72}, 095011 (2005)
  [hep-ph/0506312];
  ``Fermion mass generation in SO(10) with a unified Higgs sector,''
  Phys.\ Rev.\  D {\bf 74}, 075004 (2006),
  [arXiv:hep-ph/0607244];
  ``Variety of SO(10) GUTs with Natural Doublet-Triplet Splitting via the Missing Partner Mechanism,''
  Phys.\ Rev.\ D {\bf 85}, 075002 (2012)
  [arXiv:1112.5387 [hep-ph]];
   P.~Nath and R.~M.~Syed,
  ``Yukawa Couplings and Quark and Lepton Masses in an SO(10) Model with a
  Unified Higgs Sector,''
  Phys.\ Rev.\  D {\bf 81}, 037701 (2010).


\bibitem{Ibrahim:2010va} 
  T.~Ibrahim and P.~Nath,
  ``Large Tau and Tau Neutrino Electric Dipole Moments in Models with Vector Like Multiplets,''
  Phys.\ Rev.\ D {\bf 81}, 033007 (2010)
  [arXiv:1001.0231 [hep-ph]].

\bibitem{Ibrahim:2011im} 
  T.~Ibrahim and P.~Nath,
  ``The Chromoelectric Dipole Moment of the Top Quark in Models with Vector Like Multiplets,''
  Phys.\ Rev.\ D {\bf 84}, 015003 (2011)
  [arXiv:1104.3851 [hep-ph]].

\bibitem{Ibrahim:2010hv} 
  T.~Ibrahim and P.~Nath,
  ``The Top quark electric dipole moment in an MSSM extension with vector like multiplets,''
  Phys.\ Rev.\ D {\bf 82}, 055001 (2010)
  [arXiv:1007.0432 [hep-ph]].

\bibitem{Ibrahim:2008gg} 
  T.~Ibrahim and P.~Nath,
  ``An MSSM Extension with a Mirror Fourth Generation, Neutrino Magnetic Moments and LHC Signatures,''
  Phys.\ Rev.\ D {\bf 78}, 075013 (2008)
  [arXiv:0806.3880 [hep-ph]].

\bibitem{Ibrahim:2009uv} 
  T.~Ibrahim and P.~Nath,
 ``On the Possible Observation of Mirror Matter,''
  Nucl.\ Phys.\ Proc.\ Suppl.\  {\bf 200-202}, 161 (2010)
  [arXiv:0910.1303 [hep-ph]].

\bibitem{Babu:2008ge}
  K.~S.~Babu, I.~Gogoladze, M.~U.~Rehman and Q.~Shafi,
  ``Higgs Boson Mass, Sparticle Spectrum and Little Hierarchy Problem in
 Extended MSSM,''
  Phys.\ Rev.\  D {\bf 78}, 055017 (2008).

\bibitem{Liu:2009cc}
  C.~Liu,
  ``Supersymmetry and Vector-like Extra Generation,''
  Phys.\ Rev.\  D {\bf 80}, 035004 (2009)
  [arXiv:0907.3011 [hep-ph]].

\bibitem{Martin:2009bg}
  S.~P.~Martin,
  ``Extra vector-like matter and the lightest Higgs scalar boson mass in
  low-energy supersymmetry,''
  Phys.\ Rev.\  D {\bf 81}, 035004 (2010)
  [arXiv:0910.2732 [hep-ph]];
  ``Raising the Higgs mass with Yukawa couplings for isotriplets in vector-like
  extensions of minimal supersymmetry,''
  Phys.\ Rev.\  D {\bf 82}, 055019 (2010)
  [arXiv:1006.4186 [hep-ph]];
  ``Quirks in supersymmetry with gauge coupling unification,''
  Phys.\ Rev.\  D {\bf 83}, 035019 (2011)
  [arXiv:1012.2072 [hep-ph]].

\bibitem{Graham:2009gy}
  P.~W.~Graham, A.~Ismail, S.~Rajendran and P.~Saraswat,
  ``A Little Solution to the Little Hierarchy Problem: A Vector-like
  Generation,''
  arXiv:0910.3020 [hep-ph].

\bibitem{previous} 
R.~L.~Arnowitt and P.~Nath,
  ``mu $\to$ e gamma and tau $\to$ mu gamma decays in string models with E(6) symmetry,''
  Phys.\ Rev.\ Lett.\  {\bf 66}, 2708 (1991).


\bibitem{bhs}
   R.~Barbieri, L.~J.~Hall and A.~Strumia,
  ``Violations of lepton flavor and CP in supersymmetric unified theories,''
  Nucl.\ Phys.\ B {\bf 445}, 219 (1995)
  [hep-ph/9501334].

\bibitem{Iltan:2001rp} 
  E.~O.~Iltan,
  ``Electric dipole moments of charged leptons and lepton flavor violating interactions in the general two Higgs doublet model,''
  Phys.\ Rev.\ D {\bf 64}, 013013 (2001)
  [hep-ph/0101017].
 
\bibitem{Lavignac:2001vp} 
  S.~Lavignac, I.~Masina and C.~A.~Savoy,
  ``Tau $\to$ mu gamma and mu $\to$ e gamma as probes of neutrino mass models,''
  Phys.\ Lett.\ B {\bf 520}, 269 (2001)
  [hep-ph/0106245].
  

 
\bibitem{Cheung:2001sb} 
  K.~-m.~Cheung and O.~C.~W.~Kong,
  ``Muon $\to$ e gamma from supersymmetry without R-parity,''
  Phys.\ Rev.\ D {\bf 64}, 095007 (2001)
  [hep-ph/0101347].
 
\bibitem{Abada:2008ea} 
  A.~Abada, C.~Biggio, F.~Bonnet, M.~B.~Gavela and T.~Hambye,
  ``mu $\to$ e gamma and tau $\to$ l gamma decays in the fermion triplet seesaw model,''
  Phys.\ Rev.\ D {\bf 78}, 033007 (2008)
  [arXiv:0803.0481 [hep-ph]].
 
 
\bibitem{Davidson:2010xv} 
  S.~Davidson and G.~J.~Grenier,
  ``Lepton flavour violating Higgs and tau to mu gamma,''
  Phys.\ Rev.\ D {\bf 81}, 095016 (2010)
  [arXiv:1001.0434 [hep-ph]].
  
 
  
\bibitem{Moyotl:2012zz} 
  A.~Moyotl and G.~Tavares-Velasco,
  ``Weak properties of the tau lepton via a spin-0 unparticle,''
  Phys.\ Rev.\ D {\bf 86}, 013014 (2012)
  [arXiv:1210.1994 [hep-ph]].
    
   
 
\bibitem{Ibrahim:2007fb}
  T.~Ibrahim and P.~Nath,
  ``CP violation from standard model to strings,''
  Rev.\ Mod.\ Phys.\  {\bf 80}, 577 (2008);
  ``Phases and CP violation in SUSY,''
  arXiv:hep-ph/0210251.
    A.~Pilaftsis,
  ``CP violation in the Higgs sector of the MSSM,''
  hep-ph/9908373.
  
   
\end{thebibliography}
\end{document}